 \newlength\smallfigwidth
 \documentclass[aps,prb,twocolumn,floatfix,showpacs,amsmath,amssymb]{revtex4}

\usepackage{graphicx}
\usepackage[hypertex]{hyperref}
\begin{document}

\preprint{UFV}

\title{Conditions for free magnetic monopoles in nanoscale square
arrays of dipolar spin ice}

\author{L.\ A.\ S.\ M\'{o}l}
\email{lucasmol@ufv.br}
\author{W.\ A.\ Moura-Melo}
\email{winder@ufv.br}
\author{A.\ R.\ Pereira}
\email{apereira@ufv.br}
\affiliation{Departamento de F\'isica, Universidade Federal de
Vi\c cosa, Vi\c cosa, 36570-000, Minas Gerais, Brazil}

\begin{abstract}
We study a modified frustrated dipolar array recently proposed by
M\"{o}ller and Moessner [Phys. Rev. Lett. \textbf{96}, 237202
(2006)], which is based on an array manufactured lithographically
by Wang \emph{et al.} [Nature (London) \textbf{439}, 303 (2006)]
and consists of introducing a height offset $h$ between islands
(dipoles) pointing along the two different lattice directions. The
ground-states and excitations are studied as a function of $h$. We
have found, in qualitative agreement with the results of
M\"{o}ller and Moessner, that the ground-state changes for
$h>h_{1}$, where $h_{1}= 0.444a$ ($a$ is the lattice parameter or
distance between islands). In addition, the excitations above the
ground-state behave like magnetic poles but confined by a string,
whose tension decreases as $h$ increases, in such a way that for
$h\approx h_1$ its value is around 20 times smaller than that for
$h=0$. The system exhibits an anisotropy in the sense that the
string tension and magnetic charge depends significantly on the
directions in which the monopoles are separated. In turn, the
intensity of the magnetic charge abruptly changes when the monopoles
are separated along the direction of the longest axis of the islands.
Such a gap is attributed to the transition from the anti to the ferromagnetic
ground-state when $h=h_1$.
\end{abstract}
\pacs{75.75.-c, 75.60.Ch, 75.60.Jk}

\maketitle

\section{Introduction}

\indent  Geometrical frustration in magnetic materials occurs when
the spins are constrained by geometry in such a way that the
pairwise interaction energy cannot be simultaneously minimized for
all constituents. A special example is an exotic class of
crystalline solid known as spin ice ($Dy_{2}Ti_{2}O_{7}$,
$Ho_{2}Ti_{2}O_{7}$). Recently, Castelnovo \emph{et al.}
\cite{Castelnovo08} have proposed that these materials are the
repository of some elegant physical phenomena: for instance,
collective excitations above its frustrated ground-state
surprisingly behave as point-like objects that are the condensed
matter analogues of magnetic monopoles. Some recent experiments
\cite{Fennell09,Morris09,Bramwell09,Kadowaki09} have reported the
observation and even the measurement of the magnetic charge and
current of these monopoles in spin ice materials; in addition,
simulations also support these ideas
\cite{Jaubert09,Castelnovo10}. Besides, to turn the research of
monopoles into a proper applied science, it will be necessary to
ask if the basic ideas of dipole
fractionalization\cite{Castelnovo08,Nussinov07} that give an usual
spin-ice material its special properties can be realized in other
magnetic settings. One of the most promising candidates for
accomplishing that, is the artificial version of spin ices
recently produced by Wang \emph{et al.} \cite{Wang06}.
In this system, elongated magnetic nano-islands are regularly distributed in a
two-dimensional ($2d$) square lattice. The longest axis of the
islands alternate its orientation pointing in the direction of the
two principal axis of the lattice \cite{Wang06}. The
magnetocrystalline anisotropy of Permalloy (the magnetic material
commonly used to fabricate artificial spin ice) is effectively zero, so
that the shape anisotropy of each island forces its magnetic
moment to align along the largest axis, making the islands
effectively Ising-like.
Actually, the fabrication and study of this kind of lower
dimensional analogues of spin ice have received a lot of attention
\cite{Wang06,Moller06,Remhof08,Ke08,Zabel09,Mol09,Libal09,Moller09}.
Indeed, the ability to manipulate the constituent degrees of
freedom in condensed matter systems and their interactions is much
important towards advancing the understanding of a variety of
natural phenomena. Particularly in this context, the possibility
of observing magnetic monopoles in artificial spin ices
\cite{Mol09,Moller09} is a timely problem given that these
magnetic compounds could provide the opportunity to see them up
close and also watch them move (for example, with the aid of
magnetic force microscopy). Very recently, the direct observation
of these defects in an artificial kagome lattice was reported by
Ladak \emph{et al.} \cite{Ladak10}. However, there is a
stimulating challenging for such an observation (or not) in
artificial square lattices as pointed out in advance.

\begin{figure}
\includegraphics[angle=0.0,width=6cm]{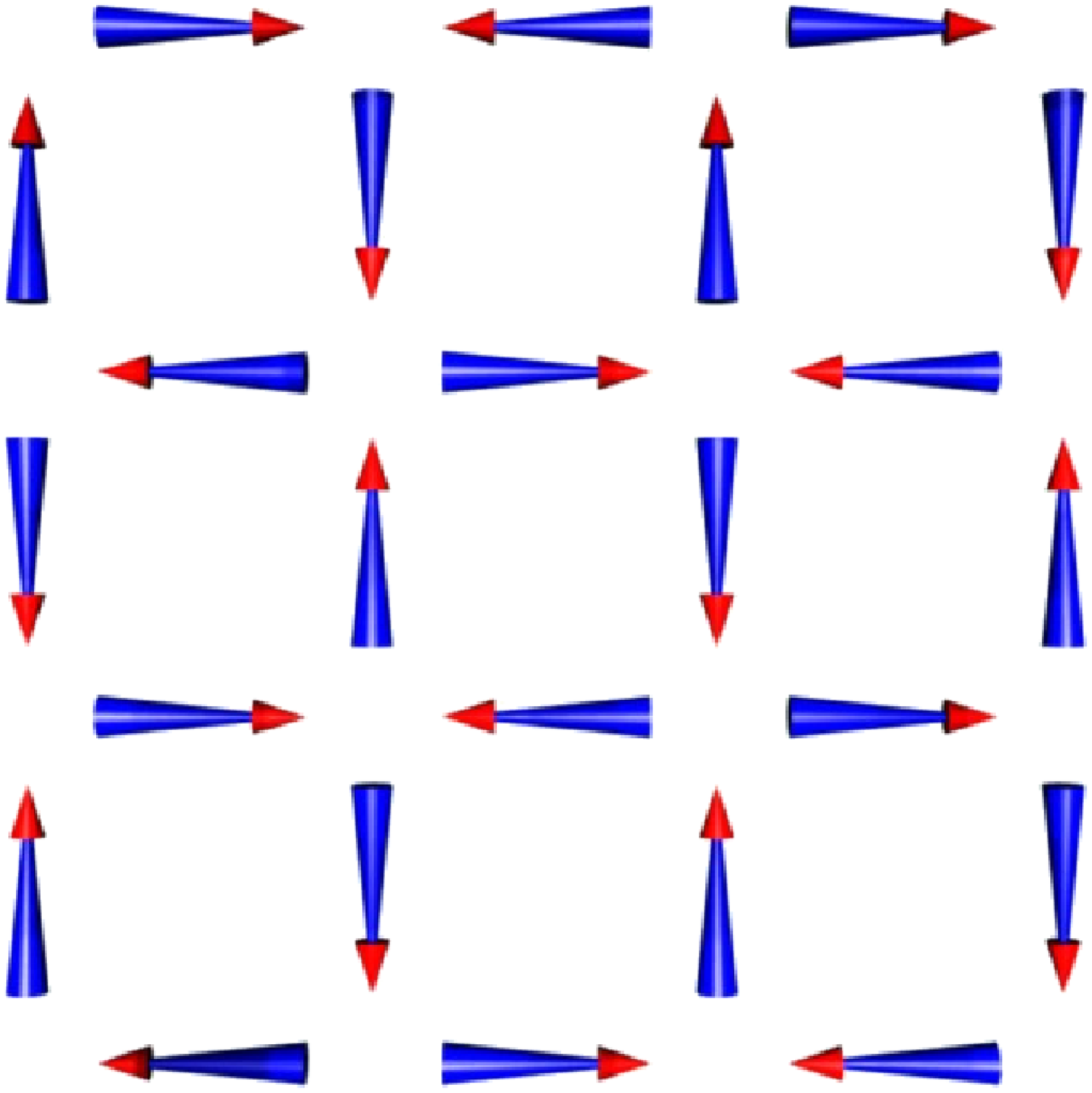}
\includegraphics[angle=0.0,width=6cm]{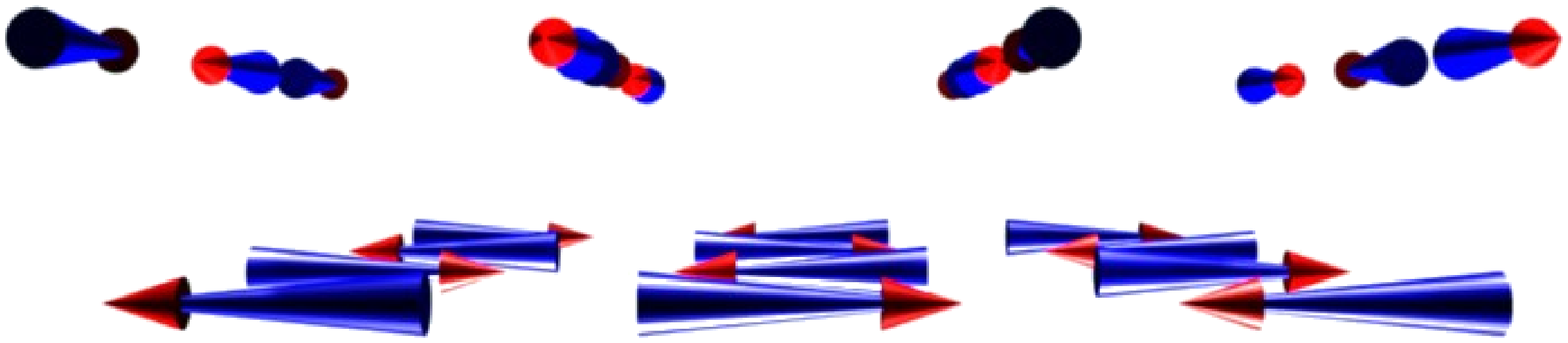}
\caption{ \label{ModifiedSystem} (Color online) The modified
square lattice studied in this work. Top: top view of the
system. The arrows represent the local dipole moments
($\vec{S}_{\alpha(i)}$ or $\vec{S}_{\beta(i)}$).
Bottom: lateral view of the system showing the height offset
between islands. The original material produced by Wang \emph{et
al.}\cite{Wang06} is two-dimensional with $h=0$.}
\end{figure}

\begin{figure}
\includegraphics[angle=0.0,width=7cm]{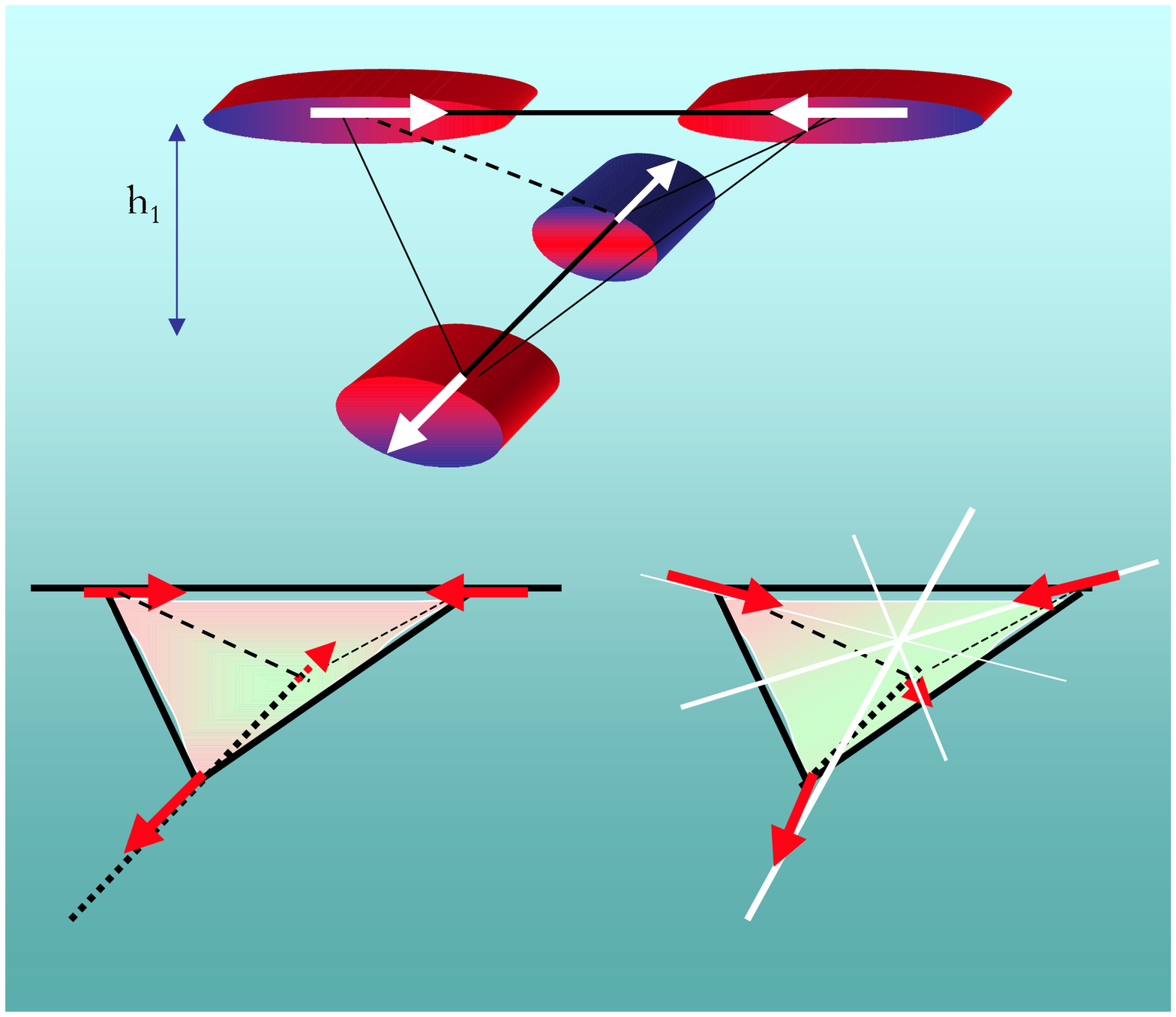}
\caption{\label{Vertices} (Color online) Up: In the artificial
spin ice proposed in Ref. ~\onlinecite{Moller06}, the spins
obeying the ice rule do not point along directions passing by the
center of a tetrahedron as they do in the natural spin ice
compounds. Down: Configurations of the spins obeying the ice rule
in a tetrahedron in the artificial (left) and the natural (right)
spin ices. This small distortion of the spins configuration causes
a residual ordering and consequently, an outstanding energetic
string connects the monopoles in the modified artificial system.}
\end{figure}

In a previous work \cite{Mol09} we have pointed out that monopoles
do not appear as effective low-energy degrees of freedom in
two-dimensional square spin ices, as they do in the
three-dimensional materials $\{Dy,Ho\}_{2}Ti_{2}O_{7}$. Due to the
antiferromagnetic order in the ground-state, the constituents of a
pair monopole-antimonopole become confined by a string which
forbids them to move independently. However, we have also argued
that above a critical temperature, the string configurational
entropy may lose its tension leaving the monopoles free. The
quantitative analysis of such a possible phase transition is under
current investigation\cite{workinprogress}. Meanwhile, other
strategies to find monopoles in synthetic spin ices have been
proposed. M\"{o}ller and Moessner \cite{Moller09} have suggested a
modification of the square lattice geometry in which they argue
that, considering a special condition, the string tension vanishes
at any temperature. This modification in the system produced by
Wang \emph{et al.}\cite{Wang06} consists of introducing a height
offset $h$ between islands pointing along the two different
directions \cite{Moller06,Moller09} (see Fig.
\ref{ModifiedSystem}; such a system is currently under
experimental planning\cite{Schiffer-priv-comm}). Their idea
comprises basically the following: if $h$ is chosen so that the
energies of all vertices obeying ice rule become degenerate, then,
an ice regime is established leaving the monopoles ``free'' to
move (indeed, there is a Coulombic interaction between the
monopoles)\cite{Moller09}. For point-like dipoles they considered
that a degenerate state is obtained when the interactions between
nearest-neighbors ($J_1$) and next-nearest-neighbors ($J_2$) are
equal, leading to the following value for the height offset where
``free'' monopoles occur: $h_{ice}\approx 0.419 a$ (where $a$ is
the lattice spacing)\cite{Moller06}. Taking into account the
finite extension of the dipoles, the height offset diminishes and
as $\epsilon \equiv 1-l/a \rightarrow 0$ ($l$ is the length of the
island), the endpoints of the islands form a tetrahedron, so that
at $h= \epsilon a/ \sqrt{2}$ the ordering disappears, and the
monopoles become free to move\cite{Moller09}.

Here we numerically calculate the energetics of the ground-states
and excitations in the modified square lattice as a function of
$h$. In our calculations we consider point-like dipoles forming
the lattice. Although the main physical aspects of  the system
must be correct with this approximation, some parameter values
(such as magnetic charge, string tension, critical height etc)
should be quantitatively altered for the realistic case in which
$l$ has a finite length. On the other hand, since we take into
account all the long-range dipole-dipole interactions, it is
expected that our results could better describe the actual system. For
instance, while in the calculations of
Refs.~\onlinecite{Moller06,Moller09}, the ground-state changes its
configuration at $h=0.419a$, our results indicate that it occurs
at $h=h_1=0.444a$. Besides, we noted that at least one of the several
configurations that satisfy the ice rule does not have the same
energy of the ``ground-states'' ($GS_1$ and $GS_2$) at this very
height, indicating that for $h=h_1$ the system is not in a completely
degenerate state.
We have also shown that the string tension
decreases rapidly as $h$ increases but it does not vanish at any
value ($h\leq a$): rather, at $h=h_1$, its strength reads about 20
times smaller than that of the usual case for $h=0$. A
possible cause of the finite strength of the string tension even
at $h_{1}$ is the fact that, concerning the spin configurations in
a tetrahedron, the artificial spin ice has a slight difference
with its natural counterpart. For the artificial compounds
proposed in Ref.~\onlinecite{Moller09}, the localized magnetic
moments forming a corner-sharing tetrahedral lattice are forced to
point along the longest axis of the islands (here, $x$- or
$y$-directions, see Fig.\ref{Vertices}) while in the original $3d$
spin ices, they point along a $<111>$ axis (indeed, in this case,
the magnetic dipoles point along axes that meet at the centers of
tetrahedra). As a result of this mismatch, there is always a
single ordered ground state in the artificial systems
, which is
responsible for the residual value of the string tension and its
anisotropy.
Another interesting result obtained here with the
point-like dipole approximation is that the magnetic charge of the
monopoles jumps as the system undergoes a transition in its ground
state. In addition, in general, this strength of the interaction
between a monopole and its antimonopole is anisotropic, depending
on the lattice direction and on the type of order. However, as
expected from the above discussions, we note that the system
anisotropy diminishes as $h$ goes to $h_{1}$. Actually, as $h$
increases from zero, the differences found in the values of the
``charges" (as distinct directions for the monopoles separation
are taken into account) decreases, and they tend to disappear as
$h\rightarrow h_{1}$, i.e., in the ice regime (nevertheless,
$h=h_{1}$ is not really an optimal ice regime, at least for
point-like dipoles).

\section{The Model and Results}

We model the system suggested in Refs.~\onlinecite{Moller06,Moller09} assuming: the magnetic
moment (``spin") of the island is replaced by a point dipole at
its center. At each site $(x_{i},y_{i},z_{i})$ of a ``square'' lattice two
spin variables are defined: $\vec{S}_{\alpha(i)}$ with components
$S_{x}=\pm 1$, $S_{y}=0, S_{z}=0$ located at
$\vec{r}_{\alpha}=(x_{i}+a/2,y_{i},h)$, and $\vec{S}_{\beta(i)}$
with components $S_{x}=0$, $S_{y}=\pm 1, S_{z}=0$ at
$\vec{r}_{\beta}=(x_{i},y_{i}+a/2,0)$. Spins pointing along the
$y$-direction and spins pointing along the $x$-direction are in
different planes, separated by a height $h$ (see Fig.~\ref{ModifiedSystem}).
Hence, in a lattice of volume $L^{2}=n^{2}a^{2}$
one gets $2\times n^{2}$ spins
(we have studied systems with $n=20,30,40,50,60,70$). Representing the spins of the islands by
$\vec{S}_{i}$, assuming either $\vec{S}_{\alpha(i)}$ or
$\vec{S}_{\beta(i)}$, then the modified artificial spin ice is
described by the following Hamiltonian
\begin{eqnarray}\label{HamiltonianSI}
H_{SI} &=& Da^{3} \sum_{i\neq j}\left[\frac{\vec{S}_{i}\cdot
\vec{S}_{j}}{r_{ij}^{3}} - \frac{3 (\vec{S}_{i}\cdot
\vec{r}_{ij})(\vec{S}_{j}\cdot \vec{r}_{ij})}{r_{ij}^{5}}\right],
\end{eqnarray}\\
where $D=\mu_{0}\mu^{2}/4\pi a^{3}$ is the coupling constant of
the dipolar interaction. The sum is performed over all
$n^2(2n^{2}-1)$ pairs of spins in the lattice for 
open boundary conditions (OBC), while for periodic boundary conditions (PBC)
a cut-off radius was introduced when $r_{ij}> n/2a$.

\begin{figure}
\includegraphics[angle=0.0,width=4.2cm]{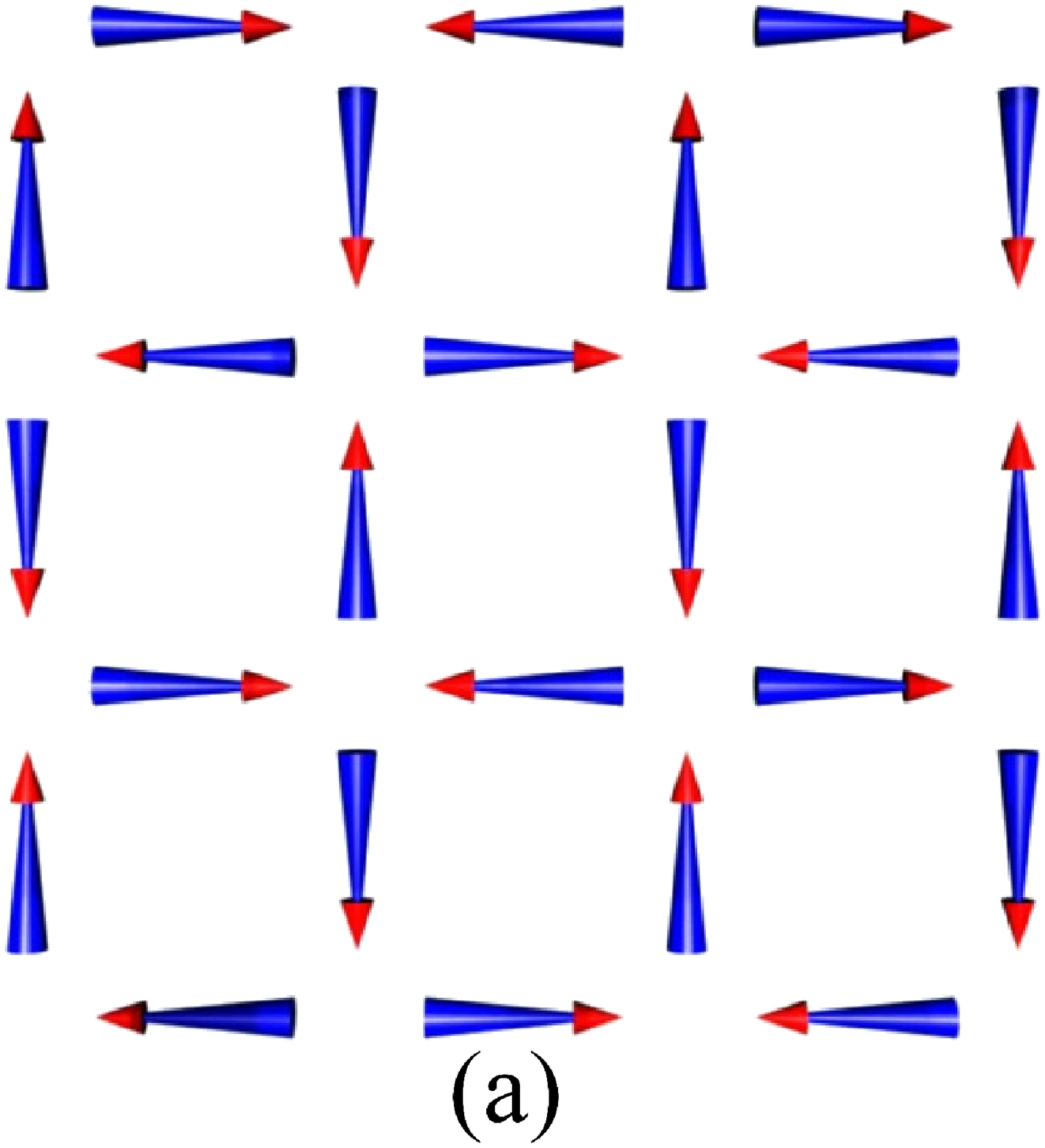}
\includegraphics[angle=0.0,width=4.2cm]{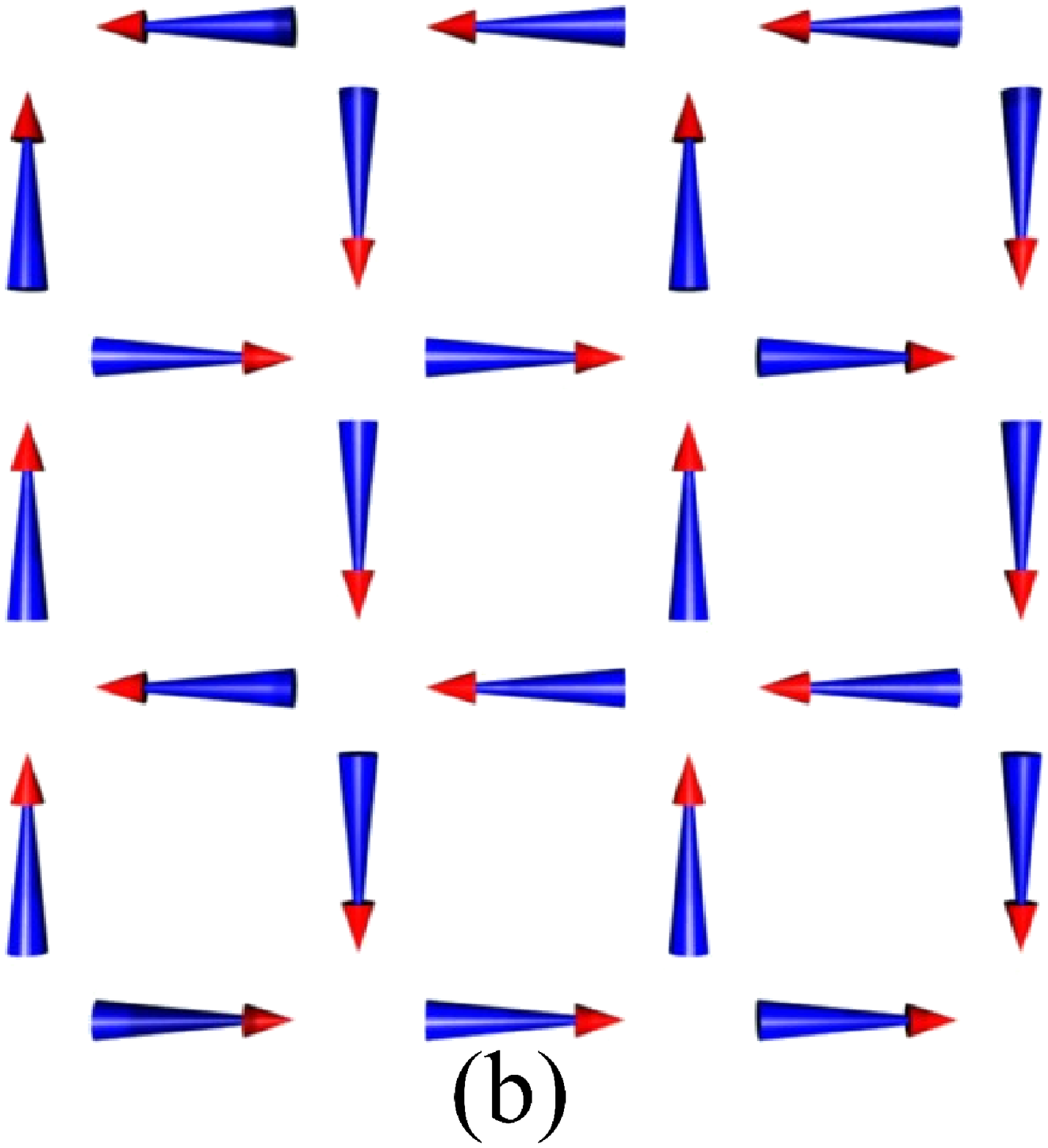}
\includegraphics[angle=0.0,width=4.2cm]{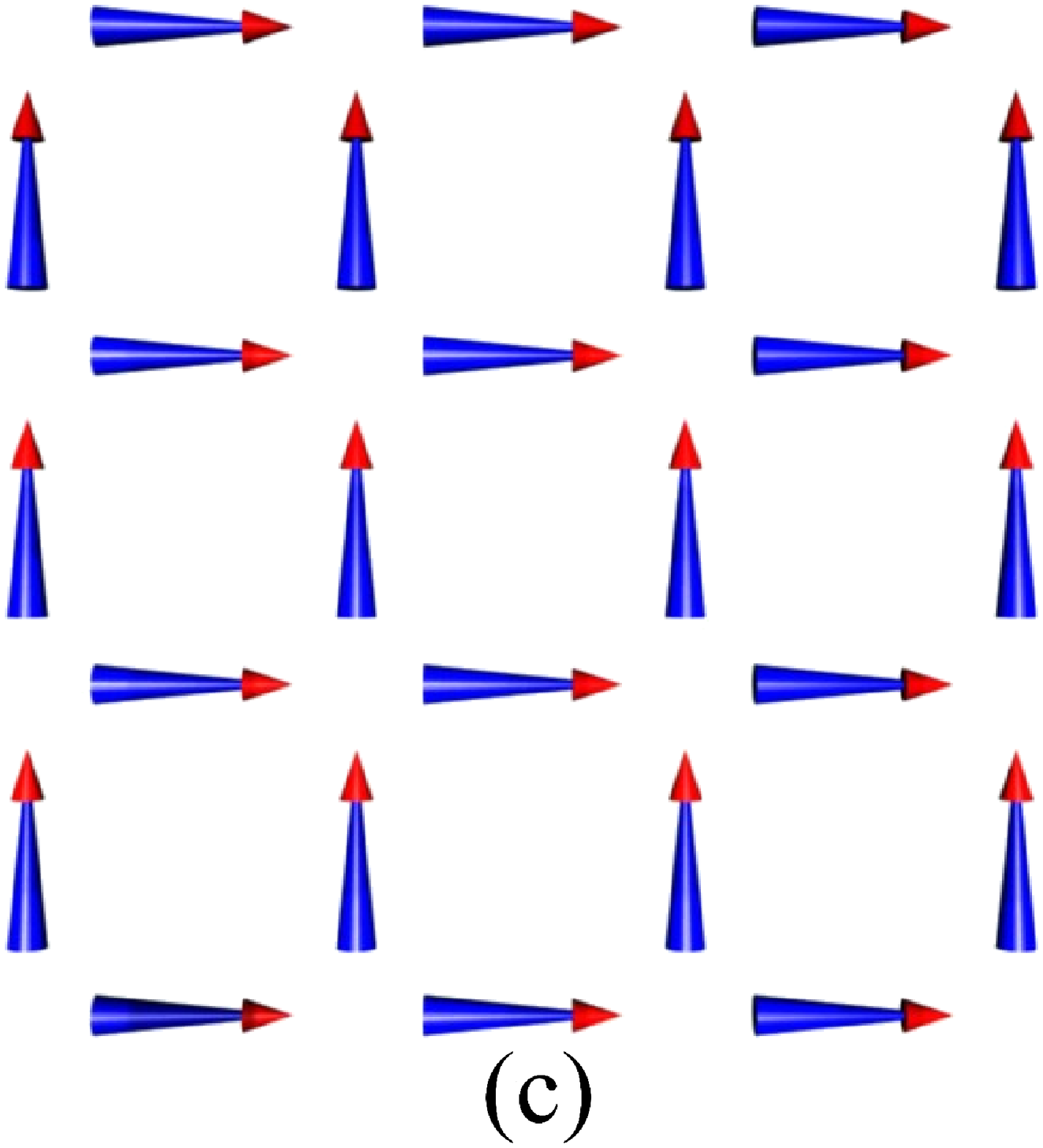}
\caption{\label{groundstates} (Color online) (a) Ground-state
configuration for $h<h_1=0.444a$, $GS_1$. Note that this is
exactly the same state obtained in
Refs.\onlinecite{Moller06,Mol09}. (b) Configuration of the
ground-state $GS_{2}$ obtained for $h>0.444$. In $GS_{2}$, each
vertex has a net magnetization but globally the magnetization
vanishes. Note that the ice rule is manifested in every vertex.
(c) Another configuration that satisfy the ice rule but has an energy
higher than the configurations shown in (a) and (b) when $h=h_1$.
}
\end{figure}

\begin{figure}
\includegraphics[angle=0.0,width=5cm]{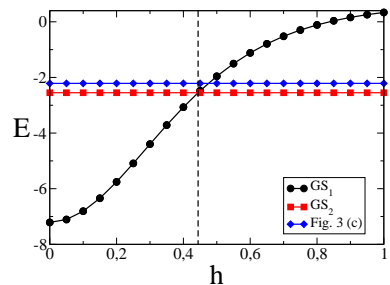}
\caption{\label{groundstatesenergy} (Color online) The energy per island of
the two ground-states ($GS_1$ and $GS_2$) and of the configuration
shown in Fig.~\ref{groundstates} (c) (in units of $D$) as a function of $h$ (in units
of the lattice spacing $a$).
Black circles represent the $GS_{1}$ energy while red squares
concern $GS_{2}$ and blue diamonds are for the configuration shown in Fig.~\ref{groundstates} (c).}
\end{figure}

The results presented here consider a lattice with $n=70$, which
contains $9800$ dipoles (islands) and PBC. We observed exactly the
same behavior for OBC and PBC and the size dependence of the
results is not appreciable. By using a simulating annealing
process (see Ref.~\onlinecite{Mol09}), the first thing to notice
is that the ground-state configuration changes for a critical
value of $h$. Indeed, as shown in
Fig.~\ref{groundstates} (a), for all values $h<h_1=0.444a$, the system
ground-state (hereafter referred to as $GS_1$)  has exactly the
same form as that of the usual case in which $h=0$. However, for
$h>h_1$, the ground-state changes to $GS_2$ (see
Fig.~\ref{groundstates} (b)). Really, as $h \to h_1$,
the energies of both states are comparable, whereas for $h>h_1$
the state $GS_2$ is less energetic (see
Fig.~\ref{groundstatesenergy}). Such a result is in qualitative
agreement with findings of Ref.~\onlinecite{Moller06}, which
presents the transition at $h=h_{ice} = 0.419a$. As expected, both
configurations obey the ice rule (two spins point in and two point
out in every vertex), but while in $GS_{1}$ the magnetization is
zero at each vertex, in $GS_{2}$ it points diagonally, but with
net vanishing magnetization. As shown in
Fig.~\ref{groundstatesenergy}, the energy of $GS_1$ increases
rapidly as $h$ increases while the energy of $GS_2$ is constant.
Actually, for this latter configuration, the horizontal and
vertical sub-lattices are decoupled. We note that these two
ground-states are metastable in the sense that they are local
minima and cannot be continuously deformed one into another
without spending a considerable amount of energy; trying to align
the dipoles from one state to another costs the inversion of two
spins by vertex (half of the spins have to be inverted in the
whole system). This changing has an $h$-dependent energy barrier
which is roughly of the order of $160D$ (for $h=0.444a$ and
$n=70$), making this process much improbable to occur
spontaneously. Thus, considering the system in the $GS_1$ state
and increasing continuously the height from $h=0$, $GS_1$ may
persist even for $h>h_{1}$ because of the large energy necessary
to change to $GS_2$. Besides, in Fig.~\ref{groundstatesenergy} we also
present the energy of the configuration shown in Fig.~\ref{groundstates} (c),
which also satisfy the ice rule but has an energy higher than those of $GS_1$
and $GS_2$ even for $h=h_1$. Consequently, the states satisfying the ice rule
are not completely degenerate.

\begin{figure}
\includegraphics[angle=0.0,width=5.0cm]{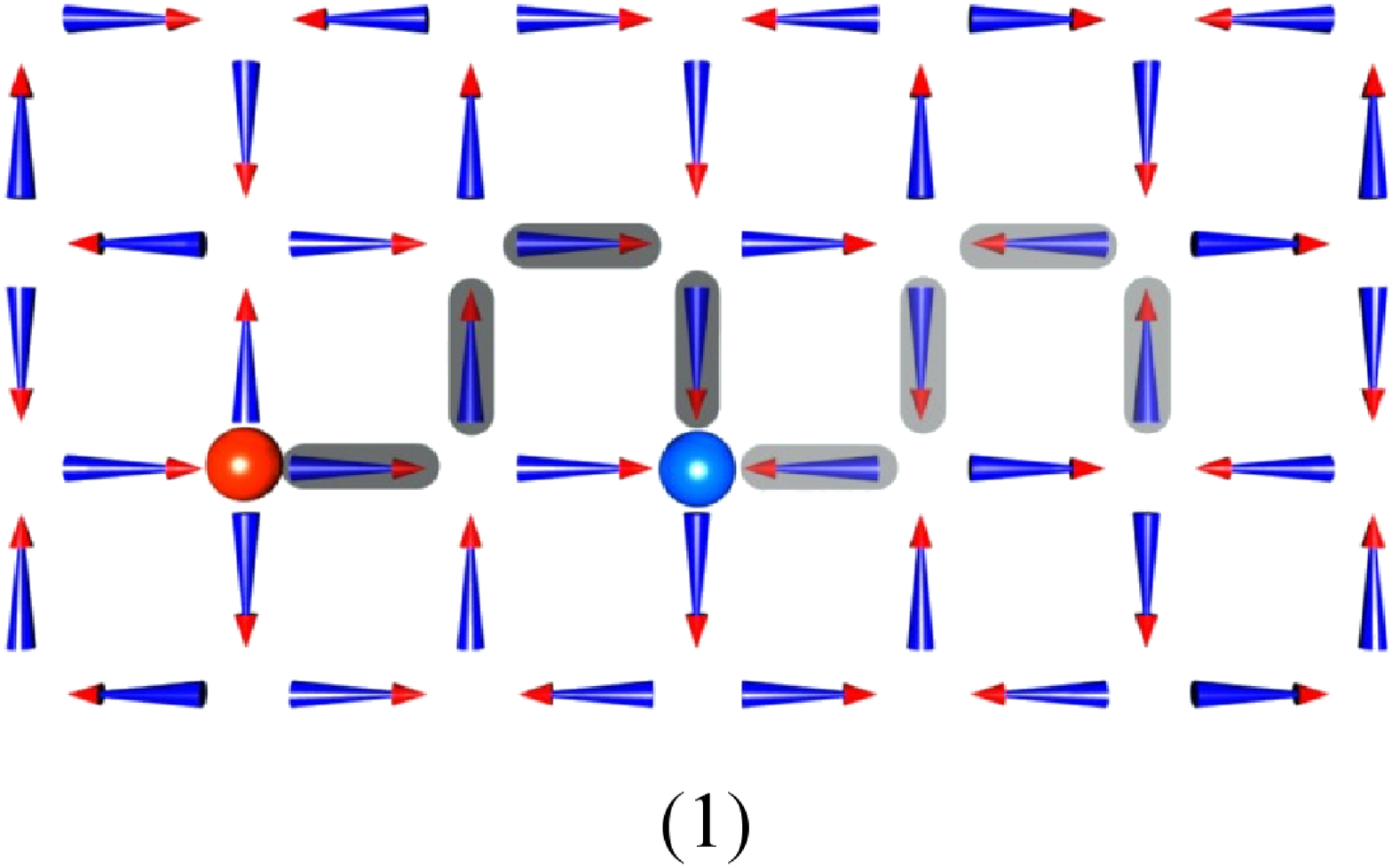}
\includegraphics[angle=0.0,width=5.0cm]{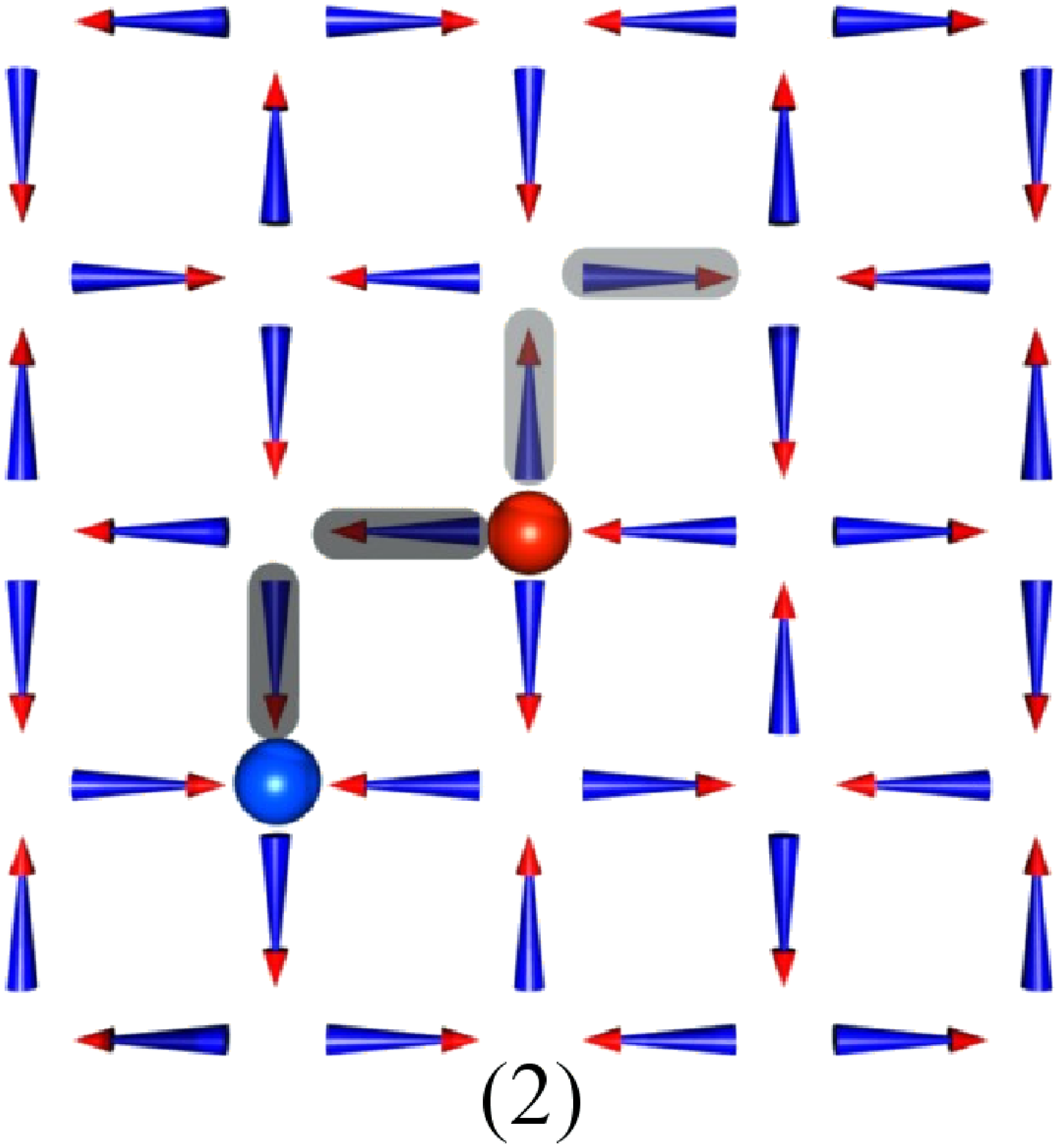}
\includegraphics[angle=0.0,width=5.0cm]{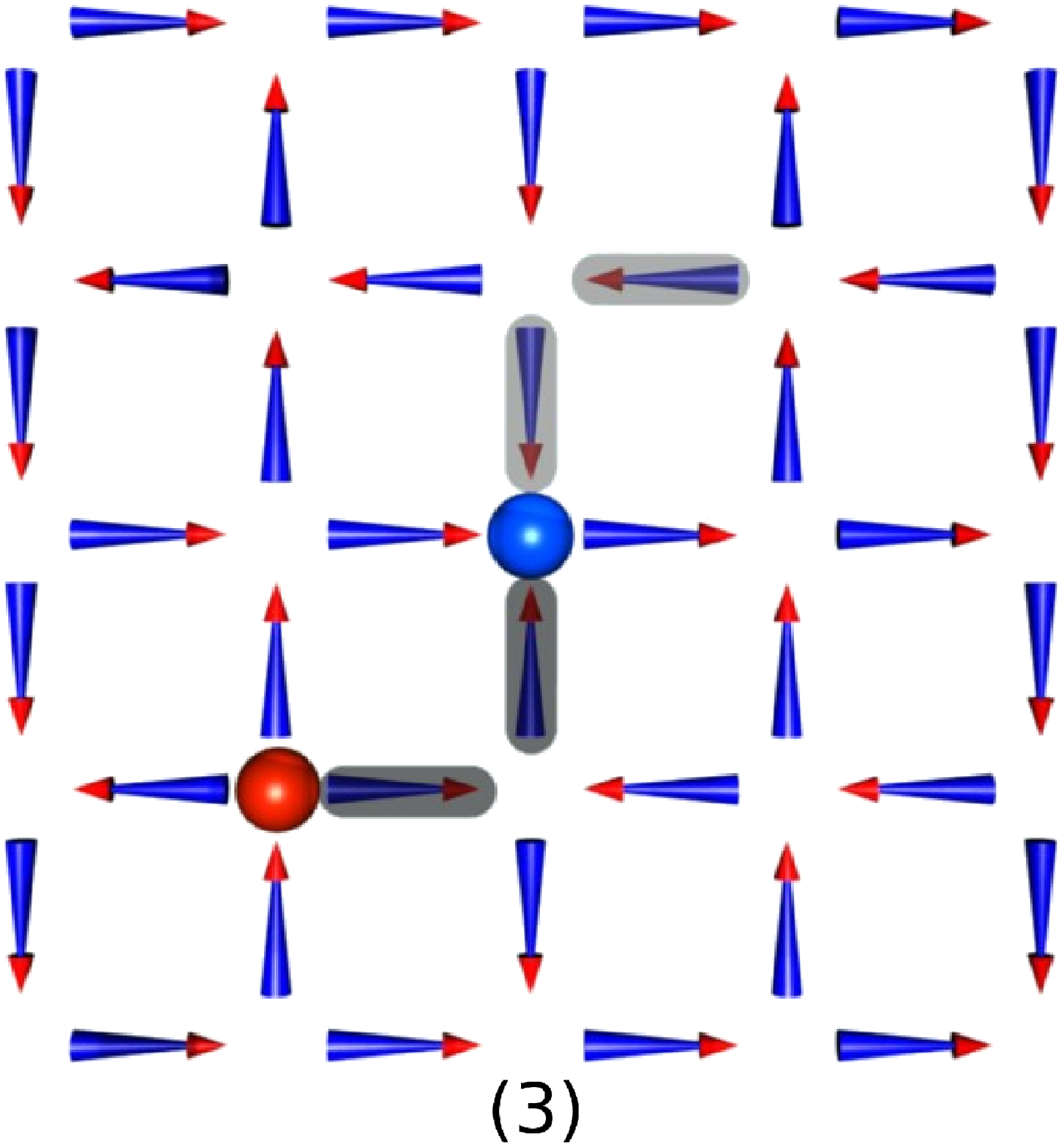}
\caption{\label{cena10} (Color online) Three of the four basic shortest
strings used in the separation process of the magnetic charges.
Pictures (1) and (2) exhibit strings $1$ and $2$,
respectively used for $h<h_1=0.444a$. The red circle is the positive charge (north pole)
while the blue circle is the negative (south pole). For $h>h_1$
the ground-state is $GS_2$ and we used a linear string-path (not
shown above) and a diagonal path (picture (3)).}
\end{figure}

\begin{figure}
\includegraphics[angle=0.0,width=5cm]{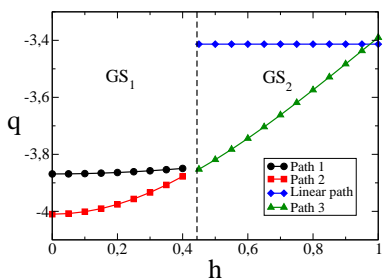}
\caption{\label{Coulomb} (Color online) The monopole ``charge''
$q$ (see Eq.~\ref{potential}) obtained analyzing the energy in the
separation process of the charges for the two string shapes shown
in Fig.\ref{cena10} for $h<h_1$. When $h>h_1$, the charges
variation is shown for a linear and diagonal string-paths. Here,
$q$ is in units of $Da$ while $h$ in units of $a$. Note how the
anisotropy of the monopole interaction decreases considerably as
$h\rightarrow h_{1}$ from below.} 
\end{figure}

Now, we consider the excitations above the ground-state. In the
two-in/two-out configuration, the effective magnetic charge
$Q_{M}^{i,j}$ (number of spins pointing inward minus the number of
spins pointing outward on each vertex $(i,j)$) is zero everywhere
for $h<h_1$ ($GS_1$) and for $h>h_1$ ($GS_2$). The most elementary
excited state is obtained by inverting a single dipole to generate
localized ``dipole magnetic charges". Such an inversion
corresponds to two adjacent sites with net magnetic charge
$Q_{M}^{i,j}=\pm 1$, which is alike a nearest-neighbor
monopole-antimonopole pair \cite{Castelnovo08,Mol09}. Following
the same method of Ref. \onlinecite{Mol09}, it is easy to observe
that such ``monopoles" can be separated from each other without
violating the local neutrality by flipping a chain of adjacent
spins. We choose four different ways they may be separated (see Fig.\ \ref{cena10}). Firstly,
using the string shape $1$ and starting in the ground-state
$GS_{1}$ (for $h<h_1=0.444a$) we choose an arbitrary site and then
the spins marked in dark gray in Fig.\ \ref{cena10} are flipped,
creating a monopole-antimonopole separated by $R=2a$. Next, the
spins marked in light gray are flipped and the separation distance
becomes $R=4a$, and so on. In this case, the string length ($X$)
is related to the charges separation distance $R$ by $X=4R/2$ (the
monopole and the antimonopole will be found along the same
horizontal or vertical line). Secondly, we also consider a
string-path of form $2$ (for $h<h_1$), making the separated
monopoles to be found in different lines (diagonally positioned;
now we have $X=2R/\sqrt{2}$). More two equivalent ways
were studied for $h>h_1$ in which $GS_2$ is the
ground-state. In this case, however, differently from the
situation in the $GS_{1}$ state, now the monopoles can be
separated by using a linear string-path (so that $X=R$) without
any violation of the ice rule. Finally, another monopoles
separation studied for $GS_{2}$ is the ``diagonal-path" (or path
3), in which the charges are put in different lines. Our analysis
shows that besides the Coulombic-type term $q(h)/R$ (where
$q=\frac{\mu_0}{4\pi}q_1q_2<0$ is the a coupling constant which
gives the strength of the interaction), the total energy cost of a
monopole-antimonopole pair has an extra contribution behaving like
$b(h)X$, brought about by the string-like excitation that binds
the monopoles, say,
\begin{equation}
V(R,h)=q(h)/R+b(h)X(R)+V_{0}(h)
\label{potential}
\end{equation}
where $V_{0}(h)$ is a $h$-depended constant related to the
monopole pair creation (for instance, for $h=0$ $V_0(0)\approx 23
$D and $V(a,0) \approx 29 $D). The results for the ``charge'' $q(h)$ are
shown in Figure \ref{Coulomb} for the range $0<h<a$. When $h<h_1$,
the excitations are considered above $GS_1$, and we observe that
there is a small h-dependent difference in the $q$-value for paths
$1$ and $2$, which vanishes as $h\to h_1$. At higher heights,
$h>h_1$, $q$ is valued with respect to $GS_2$, and is
$h$-independent for a linear string-path. However, for
path 3, it comes back to increase as $h$ increases. Therefore, the
interaction of a monopole with its partner (antimonopole) is
anisotropic in artificial spin ices. Perhaps, it would be more
appropriate to redefine things in such a way that
$q=\frac{\mu_0}{4\pi}Q_1Q_2 \alpha(h,\phi)$, where
$q_{1}q_{2}=Q_1Q_2 \alpha(h,\phi)$ and the actual value of the
charges $Q_{1}=-Q_{2}$ is independent of the angle $\phi$ that
the line connecting the poles makes with the $x$-axis. In this
case, the anisotropy of the interaction (coming from the
background) is implicitly considered in the function
$\alpha(h,\phi)$ but its complete expression was not evaluated
here. Since $\alpha(h_{1},\phi)$ tends to be a constant (independent of
$\phi$, we set $\alpha(h_{1},\phi)=1$ and so, only around the ice
regime (i.e., $h \approx h_{1}$), the interaction tends to be
isotropic. Thus we can find the genuine strength of the magnetic
charge in this artificial compound as being $ Q_1 = \pm \sqrt{4\pi
\mid q(h_{1}) \mid / \mu_{0}} \approx \pm\ 1.95 \mu/a $, where we
have used $ \mid q(h_{1}) \mid =3.8 Da $. Just for effect of
comparison, using some parameters of Ref.\onlinecite{Wang06} such
as $a=320nm$, we get a charge value which is about $80$ times
larger than the typical value found for the original $3d$ spin
ices \cite{Castelnovo08} (or about 100 times smaller than the
Dirac fundamental charge). Besides its anisotropy, another
interesting fact about the Coulombic interaction in the artificial
compounds is that it jumps at $h=h_{1}$. Indeed, at this point,
$q$ abruptly changes from $q_{<} \approx -3.8Da$ to $q_{>}\approx
-3.4Da$ when the linear path is taken into account. Such
a discontinuity may be attributed to the ground-state transition
and that above $GS_2$ the Coulombic interaction between a pair
somewhat incorporates the residual magnetization stored in each
vertex. On the other hand, keeping a diagonal separation
of the monopoles along the ground state transition, the magnetic
charge parameter $q$ increases almost continuously.

\begin{figure}
\includegraphics[angle=0.0,width=5cm]{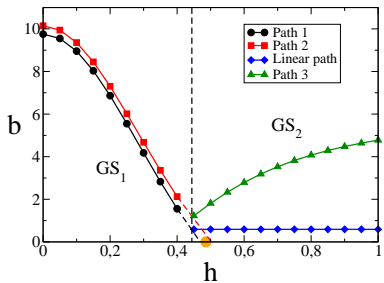}
\caption{\label{tension} (Color online) The string tension for the
two string shapes shown in Fig.\ref{cena10} for $h<h_1$ and for a
linear string-path and path 3 (diagonal) for $h>h_1$. The
green dot and the dashed lines represents an extrapolation of our
data.}
\end{figure}

How the string tension $b$ depends upon $h$ is shown in
Fig.\ref{tension}. Note that, while $GS_1$ is the
ground-state ($h<h_1$), $b$ diminishes as $h$ increases. At higher
heights, and being evaluated over $GS_2$, the tension remains a
non-vanishing small constant for linear path and turns back to
increase for diagonal separation (path 3). In general,
since $b$ is also a function of $\phi$ (i.e.,$b(h,\phi)$), it is
more favorable energetically that a pole and its antipole reside
at the same line in the array. It should be remarked
that, near the ice regime, $b(h_{1},\phi)$ is almost independent
of $\phi$ (almost isotropic limit) and its value is reduced
around 20 times whenever this modified system is compared
to its counterpart at $h=0$ (at zero temperature). In principle,
this result indicates that free monopoles do not appear in this
system. Then, the modified array \cite{Moller09} faces a small obstacle
by the fact that the islands are placed in such a way that the
spins can not point to the center of a tetrahedron as they do in
the $3d$ materials. Indeed, as pointed out before, the spins in
the artificial compound point along its edges (see
Fig.\ref{Vertices}); the islands are rigid objects that do permit
the spins to point only along their longest axis. This disparity
causes an ordering in the artificial material, which diminishes as
$h$ increases; eventually it becomes tiny but persists at
$h=h_{1}$. This persistent ordering contributes for the residual
string tension at the ice regime and also for the different string
tension values as the monopoles are located at different angular
positions in the array. Such a difficulty may be overcome when one
takes the limit $l\rightarrow a$ in the modified array. As pointed
out in Ref.\onlinecite{Moller06}, the mechanism responsible for
the equivalence between the artificial ($2d$) and natural $3d$
spin ices is not operational in $d=2$, as it requires also the
dimensionality of the dipolar interaction to coincide with that of
the underlying lattice. Here, we have a $d=3$ dipolar ($1/r^{3}$)
and Coulombic (``monopolar", $1/R$) interactions in a
two-dimensional array. Independent of this, since the state
$GS_1$ is metastable one could imagine if the excitations could be
considered to lie in $GS_1$ for $h$ slightly greater than $h_1$.
In this case the extrapolation of our results indicate that the
string tension may vanish at $h\approx 0.502 a$ (see Fig.
\ref{tension}).

\section{Summary}

In summary, we have investigated the energetics of the modified
artificial spin ice expressing several quantities, such as ground
states energy, magnetic charges and string tension, as a function
of the height offset $h$. Our analysis show that the ground-state
changes from an ordered antiferromagnetic to a ferromagnetic one
at $h=h_1\approx 0.444a$, which is in good agreement with the
value obtained in Refs.~\onlinecite{Moller06,Moller09},
$h_{ice}\approx 0.419a$. We claim that such a small difference
comes about from the fact that in these cited works, authors
assumed equal nearest-neighbor and next-nearest-neighbor
interactions, whereas we have taken all the dipole interactions
into account. For the excitations above the ground-state we have
found that the magnetic charges interact through the Coulomb
potential added by a linear confining term with tension $b(h)$,
which decreases rapidly as $h$ increases, from $0$ to $h_1$,
assuming a non-vanishing constant value at higher $h$.
Actually, the system presents an anisotropy that manifests
itself in both the Coulombic and linear interactions and it tends
to diminish as $h$ increases, almost disappearing at $h=h_{1}$.
The source of this anisotropy is a residual ordering, which still
persists even in the ice regime (at $h=h_{1}$ for point-like
dipoles). Ordering and anisotropy may disappear completely in the
ideal limit $l\rightarrow a$, $h\rightarrow 0$. Another
interesting result is that the magnetic charge jumps, depending on
the direction in which the monopoles are separated, as the system
undergoes a transition in its ground state. For a separation of
the monopoles, vertex by vertex, along the same line of vertices,
which is possible only for the $GS_{2}$ ground state, the coupling
$q$ exhibits considerably discontinuity in relation to its limit
value in the $GS_{1}$ ground state.
On the other hand, it tends to grow up continuously for the
diagonal path along the transition. Although the residual ordering
leads to a confining scenario for monopoles, its very small
strength, whenever $h\approx h_1$, signalizes a significant
tendency of monopole-pair unbinding at a critical (optimal) height
offset, even at zero-temperature. Further improvements in model
(1), for instance, taking the actual finite-size of the dipoles
into consideration, could shed some extra light to this issue.
Additionally, temperature effects may also facilitate the
conditions for free monopoles. Indeed, the string configurational
entropy is also proportional to the string size and therefore, at
a critical temperature\cite{Mol09}, on the order of $ba$, the
monopoles may become free. In view of that, for small $b$, the
monopoles should be found unbind at very low temperatures. 
As a final remark we would like to stress that these results
show that the background configuration of spins has a deep effect
in the charges interactions, being responsible for the string tension,
anisotropies and a kind of screening of the charges.

\section*{Acknowledgments}

The authors thank CNPq, FAPEMIG and CAPES (Brazilian agencies) for
financial support.

\end{document}